# Liquid-solid surface phase transformation of fluorinated fullerene on monolayer tungsten diselenide

Zhibo Song,[1,2] Qixing Wang,[1,5] Ming-Yang Li,[3,4] Lain-Jong Li,[4] Yu Jie Zheng,[1] Zhuo Wang,[1,5] Tingting Lin,[2] Dongzhi Chi,[2] Zijing Ding,*[1,5] Yu Li Huang,*[1,2] and Andrew Thye Shen Wee*[1,5]

[1]Department of Physics, National University of Singapore, 2 Science Drive 3, Singapore 117542, Singapore

[2]Institute of Materials Research & Engineering (IMRE), A*STAR (Agency for Science, Technology and Research), 2 Fusionopolis Way, Innovis, Singapore 138634, Singapore

[3]Research Center for Applied Sciences, Academia Sinica, Taipei 10617, Taiwan

[4]Physical Sciences and Engineering, King Abdullah University of Science and Technology, Thuwal 23955-6900, Saudi Arabia

[5]Centre for Advanced 2D Materials, National University of Singapore, Block S14, Level 6, 6 Science Drive 2, Singapore 117546, Singapore

**Corresponding authors**

*Email: c2ddizi@nus.edu.sg;

*Email: huangyl@imre.a-star.edu.sg;

*Email: phyweets@nus.edu.sg

## Abstract

Hybrid van der Waals (vdW) heterostructures constructed by the integration of organic molecules and two-dimensional (2D) transition metal dichalcogenide (TMD) materials have useful tunable properties for flexible electronic devices. Due to the chemically inert and atomically smooth nature of the TMD surface, well-defined crystalline organic films form atomically sharp interfaces facilitating optimal device performance. Here, the surface phase transformation of the supramolecular packing structure of fluorinated fullerene ($C_{60}F_{48}$) on single-layer (SL) tungsten diselenide ($WSe_2$) is revealed by low-temperature (LT) scanning tunnelling microscopy (STM), from thermally stable liquid to solid phases as the coverage increases. Statistical analysis of the intermolecular interaction potential reveals that the repulsive dipole-dipole interaction induced by interfacial charge transfer and substrate-mediated interactions play important roles in stabilizing the liquid $C_{60}F_{48}$ phases. Theoretical calculations further suggest that the dipole moment per $C_{60}F_{48}$ molecule varies with the surface molecule density, and the liquid-solid transformation could be understood from the perspective of the thermodynamic free energy for open systems. This study offers new insights into the growth behaviour at 2D organic/TMD hybrid heterointerfaces.

# I. INTRODUCTION

The integration of 2D materials with organic semiconductors to create layered assemblies with unique properties for flexible electronic devices has attracted increasing interest recently [1-7]. Promising device performances have been demonstrated in field-effect transistors and optoelectronics devices based on such organic-2D heterojunctions [4-6]. For instance, p-n heterojunctions have been fabricated with the integration of a p-type molecule (e.g., pentacene) and a 2D n-type semiconductor, $MoS_2$, demonstrating good photovoltaic properties [8-11]. The deposition of selected organic acceptor/donor molecules on 2D materials is also a promising non-invasive doping approach for tuning the majority charge carrier type and manipulating the transport barrier in 2D layered composites according to the application requirements [12-14]. The interfacial properties at the organic-2D heterointerfaces are complex and need to be fully understood in order to design and construct a wide range of organic-2D hybrid structures with desirable properties.

2D layered materials have atomically smooth surfaces with the absence of dangling-bonds. Organic molecules are largely bound by weak surface van der Waals (vdW) forces, and form high-quality crystalline organic films, facilitating good device performance [5, 9, 15]. The situation is complicated by the existence of multiple substrate-mediated interactions. The surfaces of 2D crystals can modulate the growth of adsorbate layers although they are chemically inert. For example, moiré patterns induced by the underlying substrate on the otherwise homogeneous 2D materials can be used as a nanoscale template to selectively adsorb atoms, molecules and nanoclusters at specific adsorption sites [16-20]. Repulsive dipole-dipole interactions induced by charge transfer (CT) are another class of interfacial interactions that have been intensively studied for different adsorbates on metallic substrates [21-24]. As a result of the subtle competition between vdW forces and multiple substrate-mediated interactions, surface phase transformations, e.g., from disordered loose-packed conformations to well-ordered close-packed structures, have been observed during the growth [22, 25-27]. Recent investigations of graphene on supported substrates revealed that the adsorbate surface reconstruction can be also tuned by CT across the heterointerfaces [12, 28, 29]. In particular, the self-assembly of organic adsorbates on layered and 2D transition metal dichalcogenides (TMDs) will be an important model system for understanding such hybrid heterointerfaces [30, 31].

In this report, we explore the phase transformation of fluorinated fullerene ($C_{60}F_{48}$) on single-layer tungsten diselenide (SL-WSe$_2$) on a graphite substrate (Fig. 1a) using low-temperature (LT) scanning tunneling microscopy (STM) supported by first-principles calculations. $C_{60}F_{48}$ is selected as a model molecule because it is a molecular acceptor [32] with applications in lithium batteries [33, 34], electronic devices [35-37] and so on [38, 39]. WSe$_2$ represents as a prototypical 2D-TMD semiconductor with a high photon quantum yield [30, 31, 40, 41]. As revealed by STM investigations, a liquid-solid surface phase transition is observed from a loose-packed liquid phase at dilute coverage to a closed-packed solid phase at high coverage. Through careful statistical analyses, we reveal that both vdW forces and multiple substrate-mediated interactions, including dipole-dipole interactions and substrate-mediated interactions, play important roles in driving the phase transformation. Theoretical calculations based on density functional theory (DFT) further confirm that the repulsive dipole-dipole interactions are induced by CT from substrate to molecules and vary with the surface molecular density. These observations offer new insights into the understanding of the interfacial interactions governing organic molecules physisorbed on 2D semiconducting TMD surfaces.

## II. EXPERIMENTAL AND CALCULATION METHODS

SL-WSe$_2$ was directly grown on a graphite substrate by CVD method using WO$_3$ and Se powders as precursor [42]. Before STM measurements, the SL-WSe$_2$ sample was degassed in a preparation chamber overnight at 400°C with base pressure better than $10^{-9}$ mbar. A ~10% overall coverage of SL-WSe$_2$ was verified by statistical analysis over the graphite surface. $C_{60}F_{48}$ (nano-C, 99.6%) were thermally deposited onto the WSe$_2$/graphite and bare graphite substrates held at room temperature. The deposition rate is about 0.04 ML per minute, where 1 ML is defined as one full close-packed monolayer $C_{60}F_{48}$. The overall coverages were further calibrated by taking statistic over large-scale STM images. After deposition, samples were transferred *in-situ* from the deposition chamber (~$10^{-9}$ mbar) to the attached STM chamber (~$10^{-10}$ mbar) for subsequent STM measurements.

All STM measurements were carried out in a high-resolution Omicron LT-STM interfaced to a Nanonis controller at 77 K, using an electrochemically etched tungsten tip under a constant current mode. Sample holder was grounded and bias voltages were applied on the tip.

The first-principles calculations were carried out using Vienna *ab initio* simulation package code [43], employing Projector augmented wave pseudopotentials [44, 45] and the Perdew-Burke-Ernzerholf form [46] of the exchange-correlation functional. To simulate different molecular coverage, supercells with different sizes were built to containing a $C_{60}F_{48}$ molecule: (4×4) $WSe_2$ on (2√7×2√7) graphene (close to the solid phase), (3√3×3√3) $WSe_2$ on (7×7) graphene, (√57×√57) $WSe_2$ on (11×11) graphene, and (6√3×6√3) $WSe_2$ on (14×14) graphene (close to the liquid phase). Here, the graphite lattice constant of 2.46 Å obtained from experimental was used, and the SL-$WSe_2$ supercells on a single-layer graphene were optimized using vdW density functional method (optB88-vdW) [47, 48]. A vacuum of 1.7 nm along z direction is included to meet periodic boundary conditions. For comparison, $C_{60}F_{48}$ molecules adsorbed on bare graphene substrate were also calculated (SI). The energy cutoff for plane waves was set to be 400 eV, and the criterion for total energy convergence was set to $10^{-4}$ eV. During geometry optimization, all atoms were relaxed until the magnitude of forces was less than 0.01 eV/Å.

## III. RESULTS AND DISCUSSION

Figure 1(a) shows a schematic structure of $C_{60}F_{48}$ adsorbed on a SL-$WSe_2$ on a graphite substrate. The SL-$WSe_2$/graphite sample was grown by chemical vapor deposition (CVD) method, and the SL-$WSe_2$ flakes can be easily distinguished by their typical triangle shapes with lateral sizes over hundreds nm on the graphite surface [30, 40]. Fig. 1(b) demonstrates a large-scale STM image (120 × 120 $nm^2$) of the SL-$WSe_2$/graphite surface with a very low coverage of $C_{60}F_{48}$ (less than 0.01 ML). Bright balls observed in the STM image are attributed to the spherical $C_{60}F_{48}$ molecules, which are exclusively adsorbed on the graphite surface (right side and left lower corner) leaving the $WSe_2$ surface empty. A line profile corresponding to the blue solid line in Fig. 1b cross over the bare $WSe_2$ surface and the $C_{60}F_{48}$ molecules is shown in the insert. It confirms a thickness of ~7 Å for the $WSe_2$ monolayer in the left side, and a height of 10 Å for a single $C_{60}F_{48}$ layer at the right side. The individual molecules are loosely adsorbed on the graphite surface with intermolecular distances of over several nm. Bare graphite surface is observable at the same time (Fig. S1 in Supporting Information, SI). Detailed analyses will be discussed later. Furthermore, the $C_{60}F_{48}$ molecules also prefer to attach to the $WSe_2$ edges due to the relatively high chemical reactivity. As the molecular coverage increases, $C_{60}F_{48}$ molecules start to adsorb in a close-packed arrangement on the WSe$_2$ surface before the graphite surface is fully covered (Fig. S1).

Figure 2 reveals the phase transformation on the SL-WSe$_2$ surface as the $C_{60}F_{48}$ coverage gradually increases from ~0.01 ML to 0.5 ML. When the overall coverage is 0.01 ML, the SL-WSe$_2$ surface is partially decorated by sparse $C_{60}F_{48}$ clusters in panel 2(a), where each cluster is composed of less than 10 molecules, and the inter-cluster spacing is larger than 5 nm. As the coverage increases slightly to 0.03 ML, small islands comprised over 10 isolated molecules are found in panel 2(b). These small islands with lateral sizes of tens of nm are identified as a "droplet-like" phase. Small molecular clusters are still observable but decrease in density, indicating no new nucleation cores form. When the coverage further increases to 0.05 ML in Fig. 2(c), large loose-packed islands extending over hundreds nm form with the merging of small $C_{60}F_{48}$ clusters and islands. In the dilute coverages (0.01, 0.03 and 0.05 ML), it is worth noting that almost every $C_{60}F_{48}$ molecule is isolated from each other with intermolecular separations much larger than its vdW diameter (~1 nm). Finally, a close-packed phase is found when the coverage is higher than 0.1 ML. Fig. 2(d) shows a solid $C_{60}F_{48}$ island coexisting with uncovered WSe$_2$ surface (left side) at 0.5 ML. From the close-up (6×6 nm$^2$) in Fig. 2(d), we can see that the molecular arrays are closely packed and well aligned. The $C_{60}F_{48}$ unit cell highlighted by a red rhombus has a lattice constant of 1.22 ± 0.04 nm, consistent with the unit cell size of fcc (111) plane of $C_{60}F_{48}$ crystal [49].

The coverage-dependent phase transformation is also reflected in the corresponding fast Fourier transform (FFT) images generated by WSxM software [50]. Bright diffuse disks are obtained in the insets of Fig. 2(a) and 2(b) due to the small disordered molecular clusters and islands respectively. A blurred ring is generated in Fig. 2(c) inset, indicating a certain distribution of intermolecular separations for the $C_{60}F_{48}$ molecules in the loose-packed molecular islands, which is a possible liquid phase. Finally, a hexagonal pattern comprised of six sharp spots is obtained in the FFT image of Fig. 2(d) inset, reflecting the formation of the close-packed well-ordered hexagonal structure above a critical coverage. A similar phase transformation is also observed on the bare graphite surface (Fig. S4).

Histograms of the intermolecular separations are demonstrated in Fig. 2(e-h). Here, only the three nearest molecule-molecule separations ($r_{int}$, from center to center) were extracted from the STM images for all coverages [22]. The technique used to extract intermolecular separations is provided in Fig. S2. Obviously, the distributions of the $C_{60}F_{48}$ molecules on the SL-WSe$_2$ surface are not random (Fig. S3) [51]. At the coverage of ~0.01 ML, a peak observed at ~1.8 nm is highlighted by a blue stripe in panel 2(e). In panel 2(f), the counts at $r_{int}$ > 2 nm increase significantly and a strong peak at ~2.8 nm (green stripe) appears, arising from the small droplet-like islands formed at 0.03 ML $C_{60}F_{48}$. The peak at ~1.8 nm (blue stripe) is still visible due to the coexisting molecular clusters, as revealed in Fig. 2(b). As the coverage increases to 0.05 ML, a significant peak at ~3.7 nm (purple stripe) is obtained in panel 2(g) corresponding to a possible liquid phase (Fig. 2(c)), while the peak at 1.8 nm almost disappears. At last, a sharp peak at 1.2 nm is observed in panel 2(h), consistent with the close-packed supramolecular structure demonstrated in Fig. 2(d). A lower cut-off at 1.2 nm (grey stripe) is observed in panels 2(e-g) for the three sparse phases, attributed to the solid phase limit.

Another method to identify the liquid or solid nature of a system is the utilization of the radial distribution function (RDF) [52] to estimate the probability of finding a molecule at a distance of $r$ away from a given reference molecule. The RDF probability is determined by

$$g(r) = \frac{1}{N^2} \sum_{i=1}^{N} \sum_{j=1}^{N} \left\langle \delta\left(\left|r_{ij}\right| - r\right) \right\rangle$$

where $r_{ij}$ is the intermolecular separation between the $i_{th}$ and $j_{th}$ molecules. Figures 3(a) and (b) provide the computed RDF curves for 0.05 ML (Fig. 2(c)) and 0.5 ML (Fig. 2(d)) $C_{60}F_{48}$ on SL-WSe$_2$ respectively after recalibration (SI). The nearest neighbor distances from more than 2000 molecules were counted for each plot. In Fig. 3(a), $g(r)$ shows a peak at 3.9 nm corresponding to the distance with the highest probability of finding a nearest neighbor molecule, which is consistent with the peak observed at the molecular distribution histogram in Fig. 2(g). The second peak appearing at 7.9 nm is twice the distance of the first peak. At larger distances, the probability approaches to a constant of around 0.065 nm$^{-2}$, which corresponds to the molecular density in the loose-packed liquid-like phase. A RDF curve for an ideal Lennard-Jones liquid on a surface is drawn as a black dashed line for comparison in Fig. 3(a). Indeed, the RDF curves obtained from our STM experiment (solid) and the ideal liquid (dashed) show agreements in the first and second coordination shells, confirming that the loose-packed conformation observed at 0.05 ML $C_{60}F_{48}$ on SL-WSe$_2$ (Fig. 2(c)) is indeed a liquid phase. The liquid $C_{60}F_{48}$ phase is less ordered than the ideal liquid, lacking long-range order as the experimental RDF curve becomes constant shortly after the second coordination shell. A minor peak at ~1.3 nm corresponding to the limit by the solid phase is observable. In Fig. 3(b), the RDF plot for the 0.5 ML $C_{60}F_{48}$ demonstrates a typical oscillating characteristic for a solid single-crystalline phase [52]. Clearly, the oscillating sharp peaks with a basic periodicity of ~1.3 nm originate from the lattice constant of the well-ordered close-packed islands shown in Fig. 2(d). At larger distance, this probability curve oscillates at around 0.7 nm$^{-2}$ corresponding to the molecular density of the solid phase $C_{60}F_{48}$ on SL-WSe$_2$. Therefore, with the combination of STM measurements and RDF plots, we have revealed the formation of $C_{60}F_{48}$ liquid and solid phases on SL-WSe$_2$ surface at different coverages with good thermal stability at 77 K.

The formation of the surface liquid phase for $C_{60}F_{48}$ on SL-WSe$_2$ is interesting due to the absence of strong molecule-substrate coupling with the chemically inert substrate at each non-polar spherical molecule. It is unlikely that the molecules are anchored at specific adsorption sites on such substrates, although moiré patterns are observable on the SL-WSe$_2$/graphite [31, 40]. The low corrugation of interaction potential in the WSe$_2$/graphite surface is one of the reasons for the observed diffusing $C_{60}F_{48}$ molecules [53-55]. Two other reasons for the observed high mobility of the molecules: tip-induced mobility and relatively weak molecule-substrate interactions which are evident by the easy desorption of molecules at low annealing temperature (100°C). In the loose-packed phases (at lower coverages), repulsive lateral interactions must be present, otherwise the molecules would aggregate into close-packed islands with the attractive vdW forces. The most likely source of the repulsive forces is the intermolecular dipole-dipole interactions induced by interfacial CT for the nonpolar $C_{60}F_{48}$ molecules, which has been observed for other atoms/molecules adsorbed on metallic substrates [56, 57]. Interfacial dipoles induced by CT has been demonstrated for solid $C_{60}F_{48}$ layer on top of a SL-WSe$_2$ in our previous study [31].

To clarify the nature of the repulsive interactions in the liquid phase, we evaluate the $C_{60}F_{48}$ intermolecular interaction potential $E(r)$ quantitatively. In terms of two-body interactions, $E(r)$ can be given by

$$E(r) = -k_B T \ln\left[ f(r) \big/ f_{ran}(r) \right]$$

where $f(r)$ is the probability distribution histogram of the nearest-neighboring (NN) molecular separations extracted from the STM image, and $f_{ran}(r)$ is the random separation distribution for noninteracting adsorbates [51, 58, 59]. The histogram of the NN molecular separations for the liquid phase is given in Fig. S3, and it is slightly different from Fig. 2(g) which was extracted from the three nearest intermolecular separations.

As demonstrated in Fig. 4, the extracted $E(r)$ (red squares) does not show a monotonic behavior, rather, local minima are observed. The valley at r = 3.5 nm with the lowest potential energy is consistent with the main peak at r = 3.7 nm in Fig. 2(c), that is, the most energetically favorable intermolecular separation. The small difference in the distance could be attributed to the different statistical method used. The local minimum at 1.4 nm gives rise to the observed limits in Fig. 2(c) and in Fig. 3(a), corresponding to the lattice constant in the well-defined solid phase. Any deviation from the minimum points would cause an increase in potential energy. Such oscillation behavior in the potential $E(r)$ cannot be explained by vdW forces only (Fig.4 blue dash curve). With the addition of repulsive dipole-dipole interactions, which decays as $1/r^3$, the fitting agrees well with the experimental $E(r)$ at small distances, but fails at large distances with a monotonic decreasing behavior (Fig. 4 pink dash curve and Fig. S9). Therefore, substrate-mediated interactions through the adsorbate-induced electronic perturbation of the surface potential should be considered [60]. That is, the surface potential can be modulated by adsorbates/defects act as scattering centers for surface electrons (e.g., Friedel oscillations [61]). Upon adsorption of another particle, the final adsorption sites of the adsorbates are determined by the (interference) interactions between the induced perturbations of both particles. Previous exploitations on the substrate-mediated interactions as a means to control the surface self-assemblies have been mainly focused on noble metals [51, 58, 59] and graphene [20, 56] substrates. Observations of Friedel oscillations induced by atomic-scale defects have been previously reported in graphene [62-65] and monolayer and bilayer $WSe_2$ [66, 67] (SI). Therefore, we can assume that the substrate-mediated interactions are significant, and the same perturbation mechanism is applicable to any substrate, whether noble metals or 2D TMD semiconductors.

Based on the theory of Hyldgaard and Persson [51, 68], the substrate-mediated potential is a function of the distance *r* (SI),

$$E_{sub} \propto -\frac{\sin(2Ar+B)}{r^2} \qquad (1)$$

where *A* is a parameter associated with the scattering wavevector.

By considering all the three components, vdW forces (Lennard-Jones potential) $E_{LJ}$, dipole-dipole interaction $E_{dipole}$, and substrate-mediated interactions $E_{sub}$ the total potential can be written as

$$E(r) = E_{LJ} + E_{dipole} + E_{sub}$$

As shown in Figure 4, the best fitting (black dashed curve) is obtained when *A* is equal to $0.6 \pm 0.2\ nm^{-1}$. The good fit to the experimental data indicates the relevance of formula (1) to the WSe$_2$ system (and possibly other 2D semiconductors). However, it is difficult to determine the momentum corresponding to parameter A, because of the multiple scattering channels in WSe$_2$ [66, 67]. Possible origins include the intravalley scattering occurring at the K point [66] and possible interfacial electronic states [69] formed at the WSe$_2$-graphite heterointerface, and further theoretical and experimental studies are required (SI).

From the final fitting (black dashed curve in Fig. 4), $\mu$ is derived to be 3.6 $e$Å for the liquid C$_{60}$F$_{48}$ on SL-WSe$_2$. DFT calculations were carried out to estimate the dipole moment $\mu$ of C$_{60}$F$_{48}$ and its origins. As discussed in the SI and our previous paper [31], the dipole originates from charge transfer from the substrate to the C$_{60}$F$_{48}$ molecule, and is of the order of several tenths $e$ per molecule for all the densities considered here. Interestingly, the charge transfer as well as the dipole moment per molecule decreases with increasing molecular density. As shown in Figure 5(a), when the molecular density is 0.1 nm$^{-2}$ (close to the liquid phase), the C$_{60}$F$_{48}$ dipole is as high as 4.4 $e$Å, which is comparable to the value of 3.6 $e$Å derived from Fig. 4. As the density increases to 0.67 nm$^{-2}$ (close to the fully close-packed monolayer), $\mu$ reduces significantly to 1.4 $e$Å per molecule. The larger molecular dipole (charge transfer) per molecule at lower density is reasonable, because each molecule probably possesses larger electrostatic interactions with the substrate when it solely occupies a larger substrate area. In addition, compared to the C$_{60}$F$_{48}$/graphite interface, C$_{60}$F$_{48}$/WSe$_2$/graphite has a relative larger molecular dipole (Fig. S12(d)). This enhancement can be explained as the molecule-graphite separation is increased by the SL-WSe$_2$ interlayer, and thus the interfacial dipole is magnified as the electron accumulation and depletion regions are mainly located in the molecule and graphite, respectively [31]. This is consistent well with our experimental fitting that gives the value of the derived dipole for the $C_{60}F_{48}$/graphite as 2.5 $e$Å (Fig. S11), which is slightly smaller. Consequently, the intermolecular separation (peak position) for the liquid $C_{60}F_{48}$ on graphite surface is also smaller (Fig. S4 and S11) indicating a higher molecular density due to the weaker intermolecular repulsive interaction.

The adsorption energy per $C_{60}F_{48}$ molecule on SL-WSe$_2$ is calculated and shown as a red curve. It is found that the adsorption energy is ~1.6 eV when the density is lower than 0.4 nm$^{-2}$, and increases to 2.25 eV at 0.67 nm$^{-2}$ (close-packing). Therefore, the molecules are less stable at sparse liquid phases although the intermolecular dipole-dipole interactions are important; the molecules have higher stability at the close-packed solid phase because the attractive vdW forces become dominant.

Finally, the phase transformation can be understood from the perspective of the thermodynamic free energy for open systems. The free energy difference $\Delta\varepsilon$ between the 2D solid and liquid surface phases is a function of the molecular coverage σ, which can be written as (the deviation is given in the SI):

$$\Delta\varepsilon \propto \Delta\gamma + \frac{\alpha}{\sqrt{\sigma}}$$

where $\Delta\gamma$ is the formation energy difference of the solid phase and liquid phase on the surface, and $\alpha > 0$ is a parameter relative to the edge formation energy (SI). As obtained from DFT calculations, $\gamma_l > \gamma_s$ and hence $\Delta\gamma < 0$. Therefore, $\Delta\varepsilon$ has a characteristic curve shown in Fig. 5b, as a reciprocal of the square root of σ. When $\Delta\varepsilon > 0$, the liquid phase is more energetically stable and thereby is preferred, otherwise the solid phase is favored. Combined with STM results, the schematic diagram is plotted in Fig. 5(b), demonstrating that the liquid phase is observed at low σ (i.e., < 0.1 ML) and the solid phase is obtained at high σ, as a result of the competition between multiple surface and interface interactions.

## IV. CONCLUSION

In summary, we have demonstrated a liquid-solid surface phase transformation of $C_{60}F_{48}$ on a SL-WSe$_2$ surface investigated by LT-STM. When the surface coverage of $C_{60}F_{48}$ molecules increases from ~0.01ML to 0.5 ML, the packing conformations transform from small molecular clusters to liquid loose-packed islands with large intermolecular separations of several nm, and finally to a solid close-packed structure. Statistical analysis of $C_{60}F_{48}$ intermolecular distances and energy potentials reveal that the repulsive dipole-dipole interactions induced by interfacial charge transfer play important roles in the formation of $C_{60}F_{48}$ liquid phases, as supported by first-principles calculations. More interestingly, the dipole moment per $C_{60}F_{48}$ molecule varies with the surface molecule density. Therefore, the phase transformation is a result of the subtle competition between vdW forces and multiple substrate-mediated interactions at different coverages, which is more complicated than expected. The present study provides new insights into the understanding of the growth behaviour of organic molecules assembling on 2D TMD surfaces, which is important for the fabrication of organic-2D hybrid heterojunctions in future flexible electronic devices.

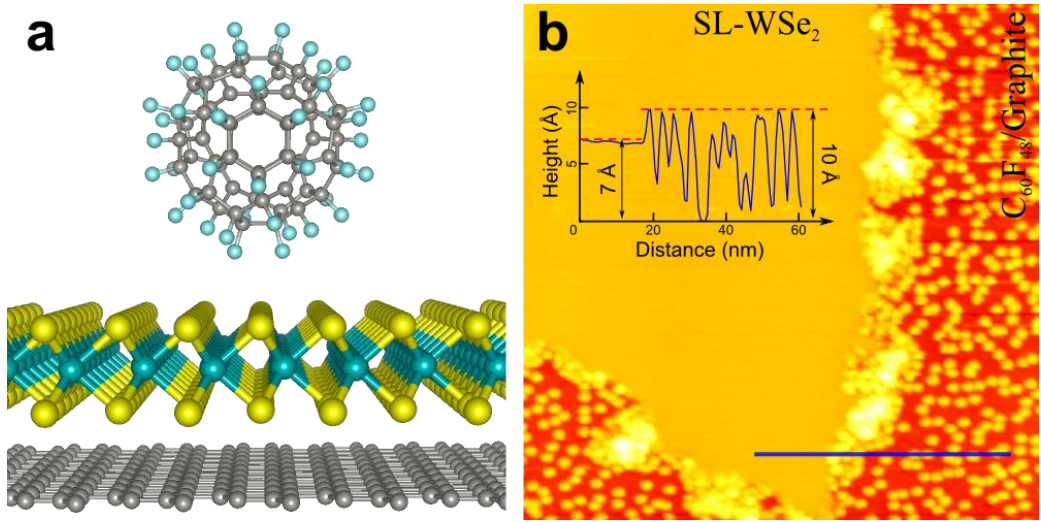

FIG 1. $C_{60}F_{48}$ molecules on SL-WSe$_2$/graphite. (a) A schematic model of the $C_{60}F_{48}$/SL-WSe$_2$/graphite heterostructure. (Grey: carbon; turquoise: fluorine; yellow: selenium; cyan: tungsten.) (b) A large-scale STM image shows a typical triangular SL-WSe$_2$ island on the graphite substrate with a very low coverage of $C_{60}F_{48}$ molecules (< 0.01 ML) ($V_{tip}$ = -3.1 V, image size: 120×120 nm$^2$). The insert lateral profile corresponding to the blue line reveals the height of 7 Å for SL-WSe$_2$ and 10 Å for $C_{60}F_{48}$ molecules.

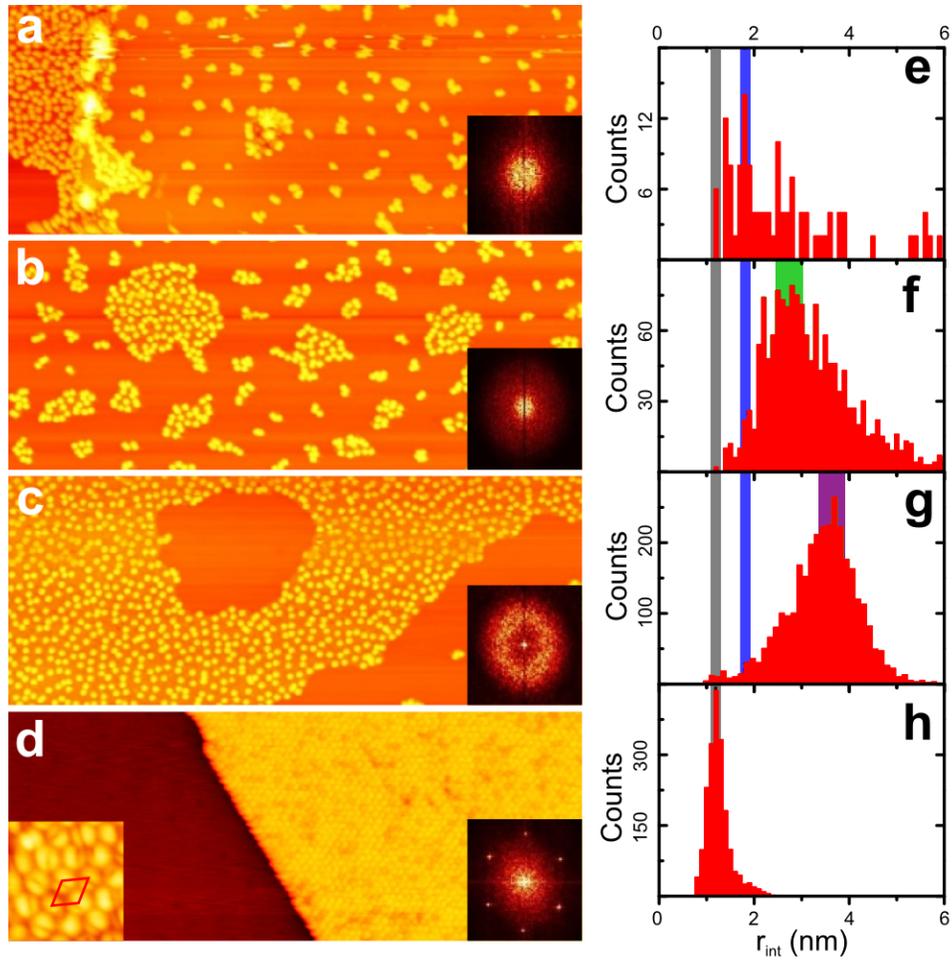

FIG 2. Liquid-solid surface phase transformation of $C_{60}F_{48}$ on SL-WSe$_2$ with the increasingcoverage. (a-d) STM images reveal the different conformations of $C_{60}F_{48}$ on SL-WSe$_2$ with the coverages of ~0.01 ML, 0.03 ML, 0.05 ML, and 0.5 ML respectively, and the inserts in the lower right corners show the corresponding FFT images. (Image size: (a-c) 86×215 nm$^2$, (d) 40×100 nm$^2$; $V_{tip}$= -3.1 V.) Left lower edge in panel (a) is $C_{60}F_{48}$ on bare graphite. The inset in the lower left corner in (d) shows a sub-molecular resolution STM image of the close-packed $C_{60}F_{48}$ island (6×6 nm$^2$; $V_{tip}$=-3.1 V). (e-h) Histograms of the three nearest molecule-molecule separations correspond to (a-d) respectively, where the colored stripes highlight the different peak positions for different coverages.

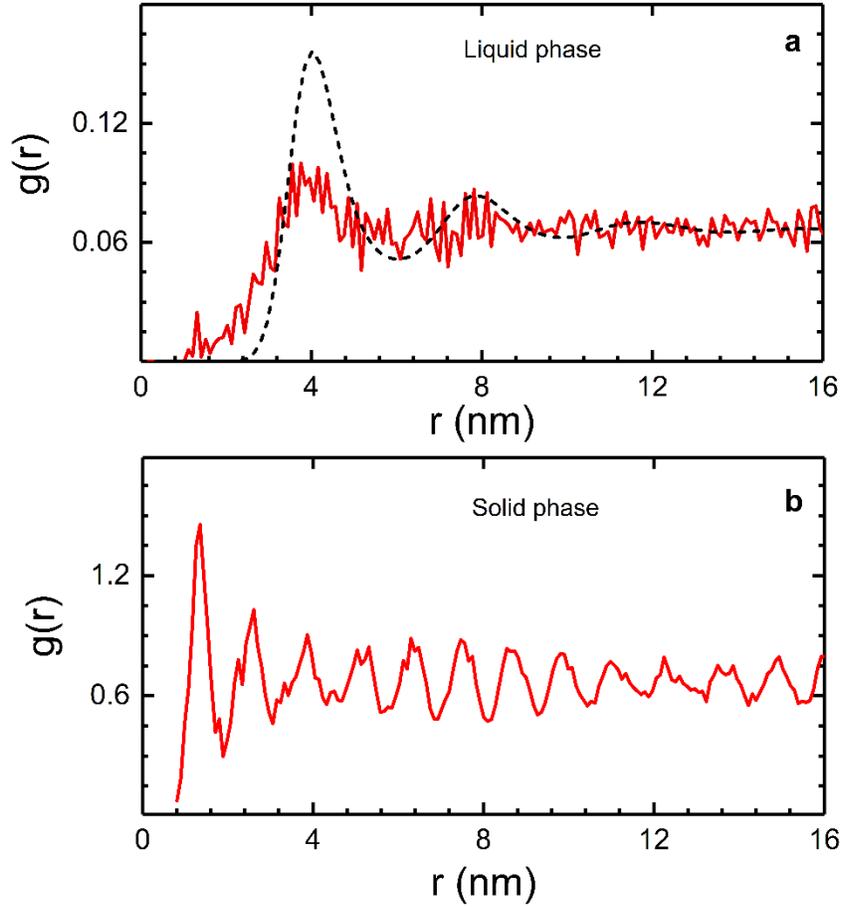

FIG 3. Radial distribution function g(r) of the liquid and solid $C_{60}F_{48}$ phases on SL-WSe$_2$. (a) Comparison of the g(r) distribution for the liquid phase obtained from STM data (red solid line) with an idea Lennard-Jones liquid (black dashed line) reveals a good agreement. g(r) is in the unit of nm$^{-2}$. (b) The g(r) distribution for the close-packed structure shows a typical oscillating characteristic for a solid phase. Here, the molecular positions and intermolecular distances have been carefully recalibrated based on the fitted eccentricity of ellipse shape FFT images and the three-fold symmetry of the WSe$_2$ lattices (SI).

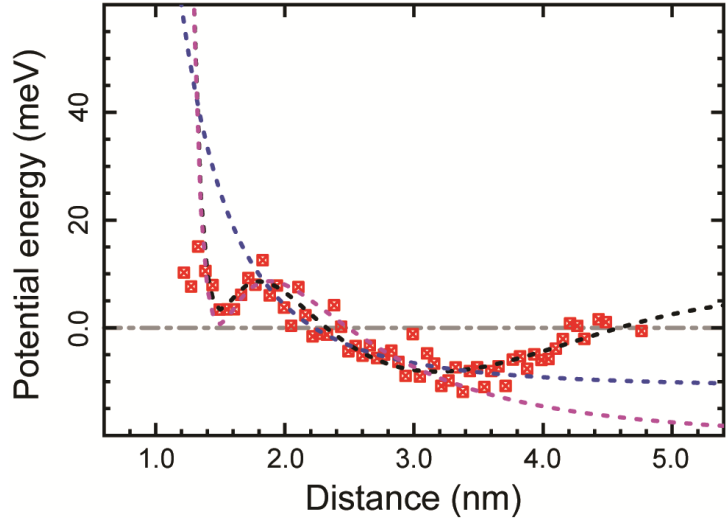

FIG 4. Evaluation of the intermolecular potential energy E(r) for $C_{60}F_{48}$ liquid phase. E(r) marked by red squares is evaluated from the histogram of the nearest-neighbour separation based on STM data. The dashed curves show the asymptotically fitting by considering: dipole-dipole interactions only (blue); dipole-dipole interactions plus vdW interaction (pink); and vdW interaction, dipole-dipole interaction and surface state mediated interactions (black). A dipole moment of 3.6 $e·Å$ for the molecule is derived for black dash curve.

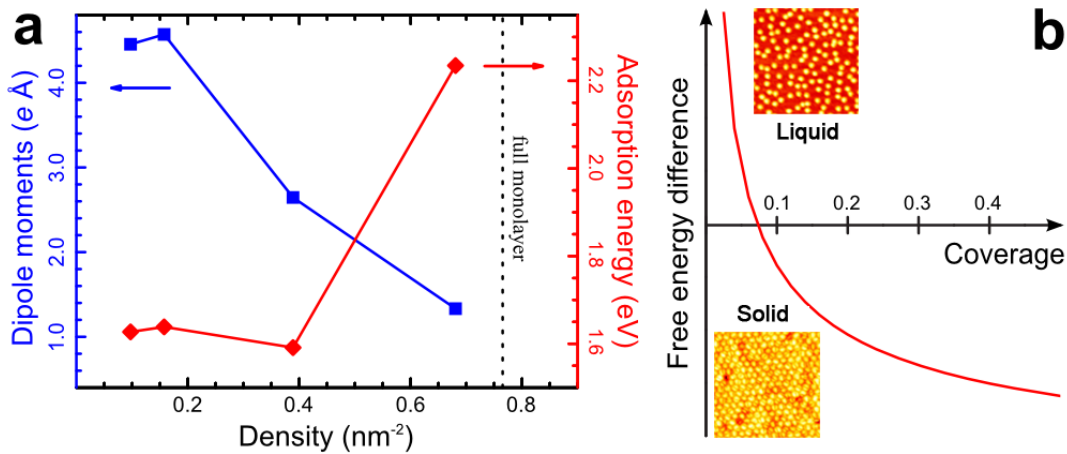

FIG 5. Theoretical models to understand the surface phase transformation. (a) DFT calculations reveal that the dipole moments per molecule of $C_{60}F_{48}$ on SL-WSe$_2$ (blue curve) decreases with the increasing molecular density, while the adsorption energy per molecule is in contrast (red curve). (b) A schematic diagram shows that the liquid phase is preferred at low coverage while the solid one is more favor at high coverage.

**SUPPORTING INFORMATION IS AVAILABLE**

**ACKNOWLEDGMENTS**

A.T.S.W. acknowledges financial support from MOE AcRF Tier 1 Grant Number R-144-000-321-112 and the Graphene Research Centre. Y.L.H. and D.C. acknowledge the A-STAR SERC grant support for the 2D growth project under the 2D pharos program (SERC 1527000012). Calculations were performed on the Graphene Research Centre cluster supported by Prof. Su Ying Quek.